\documentclass[%
print,
 amsmath,amssymb,
 aps,
]{revtex4}

\usepackage[sc]{mathpazo}
\usepackage{graphicx}
\usepackage{dcolumn}
\usepackage{bm}
\usepackage{dsfont}
\usepackage[landscape,papersize={297.1mm,210mm},left=1.9cm,right=1.6cm,top=2.7cm,bottom=2.8cm]{geometry}
\usepackage[                   
            pdfstartview=FitH,
            colorlinks, 
            pdfborder=001,   
            linkcolor=blue,
            anchorcolor=blue,
            citecolor=blue,
            urlcolor=blue
            ]{hyperref}
\usepackage{amsthm}
\makeatletter

\newcommand{\Rmnum}[1]{\expandafter\@slowromancap\romannumeral #1@}
\makeatother

\begin{document}
\title{Optimal convex approximations of quantum states based on fidelity}
\author{Huaqi Zhou$^1$}
\author{Ting Gao$^1$}
\email{gaoting@hebtu.edu.cn}
\author{Fengli Yan$^2$}
\email{flyan@hebtu.edu.cn}
\affiliation{$^1$ School of Mathematical Sciences,
 Hebei Normal University, Shijiazhuang 050024, China \\
$^2$ College of Physics, Hebei Key Laboratory of Photophysics Research and Application, Hebei Normal University, Shijiazhuang 050024, China}

\begin{abstract}
We investigate the problem of optimally approximating a desired state by the convex mixing of a set of available states. The problem is recasted as finding the optimal state with the minimum distance from target state in a convex set of usable states. Based on the fidelity, we define the optimal convex approximation of an expected state and present the complete exact solutions with respect to an arbitrary qubit state. We find that the optimal state based on fidelity is closer to the target state than the optimal state based on trace norm in many ranges. Finally, we analyze the geometrical properties of the target states which can be completely represented by a set of practicable states. Using the feature of convex combination, we express this class of target states in terms of three available states.\\

\end{abstract}


\maketitle

\section{Introduction}
In quantum information theory, the convex mixing is universal and plays an important role in the ensemble of quantum states, quantum channels, quantum measurements and quantum entanglement measures. Many entanglement measures of pure states are extended to the mixed states by using convex roof constructions in quantum entanglement theory, such as concurrence \cite{78,80,64,22}, entanglement of formation \cite{54}, geometric measure of entanglement \cite{34}, convex-roof extended negativity \cite{68}, $k$-ME concurrence \cite{86,112} and so on. Moreover, the concepts of separable and $k$-producible mixed states are defined by the convex combination of corresponding pure states in multipartite systems \cite{325,86,112,12,15,194,474,401}. The weights in a convex combination are actually the coefficients of the extremal points and may correspond to classical probabilities \cite{96}. In three-dimensional Hilbert space, the convex combination of several points represents a geometry with these points as its vertexes. In particular, the convex combination of two points expresses a line segment, and the convex combination of three points which are not on the same line shows a plane triangle.

Quantum states are very important in quantum mechanics. The so-called available states usually signify that they can be easily prepared and manipulated from the perspective of resource theory \cite{101}. However, many states are not readily obtained directly either from the aspect of experiment or from the feasibility in state preparing. In recent years, some researchers studied the problem of optimally approximating the target state by the different available states \cite{96,18,99,101}. It is similar to the issues of addressing the optimal convex approximations of quantum channels and establishing measures of the quantum resource. The optimal convex approximations of quantum channels can be redefined as looking for the least distinguishable channels according to the desired channel among the convex combination of a set of gainable channels \cite{17}. The measures of the quantum resource are embodied in quantum coherence, discord, entanglement and so on. Quantum coherence is regarded as the minimal distance of a quantum state to the set of incoherent states in the fixed reference orthogonal base \cite{113,19,v3,312}. The quantum discord can be considered as the minimal distance of a target state to classically correlated states \cite{104}. Quantum entanglement can be straightforwardly quantified by measuring the minimal difference between a given state and all separable states in quantum systems \cite{57,97,81}.

In the same way, we discuss the optimal convex approximations of quantum states. Let $\Omega$ denote the convex mixing of obtainable states. The optimal convex approximations of quantum states can essentially be viewed as calculating the minimum distance from the wished state to the convex set $\Omega$ and finding the corresponding optimal states. When the minimum distance vanishes, the target state can be completely represented by the set of available states. This is the most anticipated case. In this case, we call the target state CR state. Therefore, the optimal convex approximations of quantum states can be considered from two aspects. One is to explore the features of desired states with the minimum distance vanishing. This research is closely related to how to choose the set of available states. Generally the set of available states consists of the eigenstates of usable logic gates. In Ref. \cite{96,99}, the scholars studied the set $B_{3}$ of the eigenstates of all Pauli matrices. In Ref. \cite{18}, the set of eigenstates of any two Pauli matrices has been discussed. Recently, Liang et al. considered the set of eigenstates of arbitrary two or three real quantum logic gates \cite{101}. Other is to research the optimal convex approximation of expected state with the distance being strictly positive. This study depends on not only the set of available states but also the distance measures between quantum states. In the existing studies \cite{96,18,99,101}, they chose the distance between the states based on the trace norm. Apart from this, the geometrical properties of quantum states and quantum channels have also attracted extensive attention \cite{007,115,083,301,019,049,331}. These properties allow one to check and understand the desired traits of the states and channels.

In this paper, we define the minimum distance between a quantum state and the convex combination of a set of available states based on the fidelity \cite{41,91}. In Sec. \ref{Q1}, we provide the complete exact solution for optimal convex approximation of any qubit state in regard to the set $B_{3}$. The strengths and weaknesses of the optimal convex approximation based on the fidelity and trace norm are analyzed in Sec. \ref{Q2}. In Sec. \ref{Q3}, we find the relative volumes of the CR states (under different usable sets) as well as whole quantum states. Furthermore, we also represent the target quantum state with fewer available states by discovering the regularity in geometry. In Sec. \ref{Q4}, a summary is given. The appendices provide additional details on solution procedure and proofs.

\section{Optimal approximations of quantum states based on fidelity}\label{Q1}
Any qubit state $\rho$ can be characterized as \cite{00}

\begin{equation}\label{1}
\begin{aligned}
\rho=\frac{\mathbb{I}+\pmb{r}\cdot\pmb{\sigma}}{2},
\end{aligned}
\end{equation}
where $\mathbb{I}$ is the two dimensional identity operation, $\pmb{r}$ denotes the three dimensional real vector $(r_{x},r_{y},r_{z})$. The length $|\pmb{r}|=\sqrt{r_{x}^{2}+r_{y}^{2}+r_{z}^{2}}$ is not more than 1, $\pmb{\sigma}=(\sigma_{x},\sigma_{y},\sigma_{z})$ expresses a vector of Pauli matrices. As a matter of fact, each of the normalized three dimensional real vectors can uniquely represent a qubit quantum pure state.

The fidelity \cite{41} between two quantum states $\rho$ and $\varrho$ is a distance measure, which is defined as
\begin{equation}
\begin{aligned}
F(\rho,\varrho)=[\textup{tr}\sqrt{\rho^{1/2}\varrho\rho^{1/2}}]^{2}.
\end{aligned}
\end{equation}
The range of $F(\rho,\varrho)$ is from 0 to 1, if and only if two states $\rho$ and $\varrho$ are same, the fidelity is equal to 1. Suppose that two real vectors $\pmb{r}=(r_{x},r_{y},r_{z})$ and $\pmb{s}=(s_{x},s_{y},s_{z})$ satisfy the condition that length is not more than 1. Let $\pmb{r}$ and $\pmb{s}$ correspond to $\rho$ and $\varrho$ respectively. Then the fidelity between states $\rho$ and $\varrho$ has an elegant form \cite{41,91}
\begin{equation}
\begin{aligned}
F(\rho,\varrho)=\frac{1}{2}[1+\pmb{r}\cdot\pmb{s}+\sqrt{(1-|\pmb{r}|^{2})(1-|\pmb{s}|^{2})}].
\end{aligned}
\end{equation}

Given a set $B=\{|\psi_{i}\rangle,i=1,2,\ldots,n\}$, where $|\psi_{i}\rangle$ are available pure states. Based on the fidelity, we introduce the following definition.

Definition $1$. For the desired state $\rho$ and the set $B$, the optimal convex approximation is defined as
\begin{equation}
\begin{aligned}
D_{B}^{F}(\rho)=1-\textup{max}F(\rho,\sum_{i}p_{i}\rho_{i}),
\end{aligned}
\end{equation}
where $\rho_{i}=|\psi_{i}\rangle\langle\psi_{i}|$, the maximum is taken over all possible probability distributions $\{p_{i}\}$ with $p_{i}\geq 0$ and $\sum_{i}p_{i}=1$ for $i=1,2,\ldots,n$. When the probability distribution $\{p_{i}\}$ makes the fidelity between $\rho$ and $\sum_{i}p_{i}\rho_{i}$ reaching maximum, the state $\rho^{\textup{opt}}=\sum_{i}p_{i}\rho_{i}$ is called optimal state. The optimal state of a target state may not be unique.

We discuss this optimal convex approximation in terms of measure, the distance between the target state and the convex combination of the available states. Naturally, the problem of optimally approximating the target state from other aspects can also be considered, for example, the coherence of quantum states. In this case, our definition just needs to be changed appropriately by referring to any other figure of merit that quantifies the coherence of quantum states.

Now, we concretely compute the optimal convex approximation of arbitrary qubit state with regard to the set \cite{99}
\begin{equation}
\begin{aligned}
B_{3}=\{|0\rangle,|1\rangle,|2\rangle=\frac{1}{\sqrt{2}}(|0\rangle+|1\rangle),|3\rangle=\frac{1}{\sqrt{2}}(|0\rangle-|1\rangle),|4\rangle=\frac{1}{\sqrt{2}}
(|0\rangle+\textup{i}|1\rangle),|5\rangle=\frac{1}{\sqrt{2}}(|0\rangle-\textup{i}|1\rangle)\},
\end{aligned}
\end{equation}
which consists of the eigenstates of Pauli matrixes $\sigma_{x}$, $\sigma_{y}$, $\sigma_{z}$. Let $\rho_{i}=|i\rangle\langle i|$, $i=0,\cdots,5$. The convex combination $\sum_{i}p_{i}\rho_{i}$ of these states can be described by $\pmb{v}=(p_{2}-p_{3},p_{4}-p_{5},p_{0}-p_{1})$. Due to the symmetry, we only address the optimal convex approximations of qubit states in the range $r_{x},r_{y},r_{z}\in [0,1]$.

Computing the maximum of fidelity between $\rho$ and $\sum_{i}p_{i}\rho_{i}$ is an optimization problem. When the function satisfies inequality and equality conditions, we can use the Karush-Kuhn-Tucker (KKT) theorem \cite{101,10}. Consider the minimum value of the function $f(x)$. Suppose that $g_{i}(x)\leq 0$ and $h_{j}(x)=0$ are inequality constraints and equality conditions respectively, $i=1,2,\ldots,m$; $j=1,2,\ldots,k$. A function can be constructed as $G(x,\lambda_{j})=f(x)+\sum_{i}u_{i}g_{i}(x)+\sum_{j}\lambda_{j}h_{j}(x)$. The optimal solution $x^{\ast}$ of the function $f(x)$ (i.e. the local minimum point of the function $f(x)$) must satisfy the following conditions. First, inequality constraints $g_{i}(x^{\ast})\leq 0$ and equality conditions $h_{j}(x^{\ast})=0$. Second, $\nabla G(x^{\ast})=0$, where $\nabla$ is gradient operator. Third, inequality constraints $u_{i}\geq 0$, $u_{i}g_{i}(x^{\ast})=0$. If $f(x)$ and $g_{i}(x)$ are convex, and $h_{j}(x)$ are linear, the point satisfying above constraints and conditions is the optimal solution $x^{\ast}$ \cite{06}.

Obviously, for the probability distribution $\{p_{i}\}_{i=0}^{5}$, one has the inequality constraint $p_{i}\geq 0$ and the equality condition $\sum_{i}p_{i}=1$. Therefore, the function can be constructed as
\begin{equation}\label{4}
\begin{aligned}
G(p_{i},\lambda)&=-F(\rho,\sum_{i}p_{i}\rho_{i})-\sum_{i}\lambda_{i}p_{i}-\lambda(\sum_{j}p_{j}-1)\\
 &=-\frac{1}{2}\{1+r_{x}(p_{2}-p_{3})+r_{y}(p_{4}-p_{5})+r_{z}(p_{0}-p_{1})+\sqrt{(1-|\pmb{r}|^{2})[1-(p_{2}-p_{3})^{2}-(p_{4}-p_{5})^{2}-(p_{0}-p_{1})^{2}]}\}\\
 &\quad -\sum_{i}\lambda_{i}p_{i}-\lambda(\sum_{j}p_{j}-1),
\end{aligned}
\end{equation}
where $\lambda_{i}\geq 0$. According to the three conditions above, the optimization problem is equivalent to solving the following equations
\begin{subequations}\label{2}
\begin{align}
&\frac{\partial G}{\partial p_{0}}=\frac{(p_{0}-p_{1})\sqrt{1-|\pmb{r}|^{2}}}{2\sqrt{1-(p_{2}-p_{3})^{2}-(p_{4}-p_{5})^{2}-
(p_{0}-p_{1})^{2}}}-\frac{r_{z}}{2}-\lambda_{0}-\lambda=0,\\
&\frac{\partial G}{\partial p_{1}}=-\frac{(p_{0}-p_{1})\sqrt{1-|\pmb{r}|^{2}}}{2\sqrt{1-(p_{2}-p_{3})^{2}-(p_{4}-p_{5})^{2}-
(p_{0}-p_{1})^{2}}}+\frac{r_{z}}{2}-\lambda_{1}-\lambda=0,\\
&\frac{\partial G}{\partial p_{2}}=\frac{(p_{2}-p_{3})\sqrt{1-|\pmb{r}|^{2}}}{2\sqrt{1-(p_{2}-p_{3})^{2}-(p_{4}-p_{5})^{2}-
(p_{0}-p_{1})^{2}}}-\frac{r_{x}}{2}-\lambda_{2}-\lambda=0,\\
&\frac{\partial G}{\partial p_{3}}=-\frac{(p_{2}-p_{3})\sqrt{1-|\pmb{r}|^{2}}}{2\sqrt{1-(p_{2}-p_{3})^{2}-(p_{4}-p_{5})^{2}-
(p_{0}-p_{1})^{2}}}+\frac{r_{x}}{2}-\lambda_{3}-\lambda=0,\\
&\frac{\partial G}{\partial p_{4}}=\frac{(p_{4}-p_{5})\sqrt{1-|\pmb{r}|^{2}}}{2\sqrt{1-(p_{2}-p_{3})^{2}-(p_{4}-p_{5})^{2}-
(p_{0}-p_{1})^{2}}}-\frac{r_{y}}{2}-\lambda_{4}-\lambda=0,\\
&\frac{\partial G}{\partial p_{5}}=-\frac{(p_{4}-p_{5})\sqrt{1-|\pmb{r}|^{2}}}{2\sqrt{1-(p_{2}-p_{3})^{2}-(p_{4}-p_{5})^{2}-
(p_{0}-p_{1})^{2}}}+\frac{r_{y}}{2}-\lambda_{5}-\lambda=0,\\
&\frac{\partial G}{\partial \lambda}=\sum_{j}p_{j}-1=0,~\lambda_{i}p_{i}=0,~\lambda_{i}\geq 0,~p_{i}\geq 0,~i=0,\cdots,5.
\end{align}
\end{subequations}
Next we will show the exact solution of the equation ($\ref{2}$), for the detailed procedure please refer to the Appendix \ref{A}.

(V1) In the set $S1=\{(r_{x},r_{y},r_{z})|r_{x}+r_{y}+r_{z}\leq 1\}$, $D_{B_{3}}^{F}(\rho)=0$, which means that the target state $\rho$ can be completely represented by the convex combination of $\{\rho_{i}\}$. The corresponding coefficients are given by
\begin{equation}
\begin{aligned}
&p_{0}=\frac{1}{2}-\frac{r_{x}}{2}-\frac{r_{y}}{2}+\frac{r_{z}}{2}-t_{1}-t_{2},\\
&p_{1}=\frac{1}{2}-\frac{r_{x}}{2}-\frac{r_{y}}{2}-\frac{r_{z}}{2}-t_{1}-t_{2},\\
&p_{2}=r_{x}+t_{2},\\
&p_{3}=t_{2},\\
&p_{4}=r_{y}+t_{1},\\
&p_{5}=t_{1},
\end{aligned}
\end{equation}
where $t_{1}$ and $t_{2}$ are arbitrary non-negative numbers such that $p_{1}\geq 0$.

(V2) In the set $S2=\{(r_{x},r_{y},r_{z})|r_{x}+r_{y}< \sqrt{1-2r_{z}^{2}}+r_{z}~\textup{or}~r_{x}+r_{y}< 2r_{z},r_{y}+r_{z}< \sqrt{1-2r_{x}^{2}}+r_{x}~\textup{or}~r_{y}+r_{z}< 2r_{x},r_{x}+r_{z}< \sqrt{1-2r_{y}^{2}}+r_{y}~\textup{or}~r_{x}+r_{z}< 2r_{y}\}\setminus S1$,
\begin{equation}
\begin{aligned}
D_{B_{3}}^{F}(\rho)=\frac{1}{2}-\frac{r}{6}-\frac{\sqrt{2(3-r^{2})}}{6},
\end{aligned}
\end{equation}
where $r=r_{x}+r_{y}+r_{z}$, with the optimal weights
\begin{equation}\label{3}
\begin{aligned}
&p_{0}=\frac{1}{3}-\frac{\sqrt{2}(r_{x}+r_{y}-2r_{z})}{3\sqrt{3-(r_{x}+r_{y}+r_{z})^{2}}},\\
&p_{2}=\frac{1}{3}-\frac{\sqrt{2}(-2r_{x}+r_{y}+r_{z})}{3\sqrt{3-(r_{x}+r_{y}+r_{z})^{2}}},\\
&p_{4}=\frac{1}{3}-\frac{\sqrt{2}(r_{x}-2r_{y}+r_{z})}{3\sqrt{3-(r_{x}+r_{y}+r_{z})^{2}}},\\
&p_{1}=p_{3}=p_{5}=0.
\end{aligned}
\end{equation}

(V3) In the set $S3=\{(r_{x},r_{y},r_{z})|r_{x}+r_{z}>\sqrt{1-r_{y}^{2}}\}\setminus S1\cup S2$,
\begin{equation}
\begin{aligned}
D_{B_{3}}^{F}(\rho)=\frac{1}{2}-\frac{\sqrt{2-(r_{x}+r_{z})^{2}-2r_{y}^{2}}+(r_{x}+r_{z})}{4},
\end{aligned}
\end{equation}
with the pertaining optimal coefficients
\begin{equation}
\begin{aligned}
&p_{0}=\frac{1}{2}-\frac{r_{x}-r_{z}}{2\sqrt{2-(r_{x}+r_{z})^{2}-2r_{y}^{2}}},\\
&p_{2}=\frac{1}{2}-\frac{-r_{x}+r_{z}}{2\sqrt{2-(r_{x}+r_{z})^{2}-2r_{y}^{2}}},\\
&p_{1}=p_{3}=p_{4}=p_{5}=0.
\end{aligned}
\end{equation}

(V4) In the set $S4=\{(r_{x},r_{y},r_{z})|r_{y}+r_{z}>\sqrt{1-r_{x}^{2}}\}\setminus S1\cup S2$,
\begin{equation}
\begin{aligned}
D_{B_{3}}^{F}(\rho)=\frac{1}{2}-\frac{\sqrt{2-(r_{y}+r_{z})^{2}-2r_{x}^{2}}+(r_{y}+r_{z})}{4},
\end{aligned}
\end{equation}
with the corresponding weights
\begin{equation}
\begin{aligned}
&p_{0}=\frac{1}{2}-\frac{r_{y}-r_{z}}{2\sqrt{2-(r_{y}+r_{z})^{2}-2r_{x}^{2}}},\\
&p_{4}=\frac{1}{2}-\frac{-r_{y}+r_{z}}{2\sqrt{2-(r_{y}+r_{z})^{2}-2r_{x}^{2}}},\\
&p_{1}=p_{2}=p_{3}=p_{5}=0.
\end{aligned}
\end{equation}

(V5) In the set $S5=\{(r_{x},r_{y},r_{z})|r_{x}+r_{y}>\sqrt{1-r_{z}^{2}}\}\setminus S1\cup S2$,
\begin{equation}
\begin{aligned}
D_{B_{3}}^{F}(\rho)=\frac{1}{2}-\frac{\sqrt{2-(r_{x}+r_{y})^{2}-2r_{z}^{2}}+(r_{x}+r_{y})}{4},
\end{aligned}
\end{equation}
the related optimal coefficients are given by
\begin{equation}
\begin{aligned}
&p_{2}=\frac{1}{2}-\frac{r_{y}-r_{x}}{2\sqrt{2-(r_{y}+r_{x})^{2}-2r_{z}^{2}}},\\
&p_{4}=\frac{1}{2}-\frac{-r_{y}+r_{x}}{2\sqrt{2-(r_{y}+r_{x})^{2}-2r_{z}^{2}}},\\
&p_{0}=p_{1}=p_{3}=p_{5}=0.
\end{aligned}
\end{equation}

It needs to notice that there may be intersections between the sets $S3$, $S4$ and $S5$. If a quantum state $(r_{x}^{0},r_{y}^{0},r_{z}^{0})$ belongs to all three sets, then the least optimal convex approximation in the three cases is the genuine optimal solution.

Clearly, we have obtained the optimal solution for arbitrary qubit state. Specially, in the set $S1$, the distance between target state and the convex combinations of the available states in $B_{3}$ vanishes, which is the most desired case. These target states are all CR states. Further, we will study their geometric property in Sec. \ref{Q3}. In the cases of three numbers of $\{p_{i}\}$ being nonzero, the optimal convex approximation has a solution only if $p_{0},p_{2},p_{4}\neq 0$. The value range of this solution is the set $S2$. In geometry, it is not difficult to find that the quantum state which cannot be completely represented is located at above the plane consisting of $|0\rangle$, $|2\rangle$ and $|4\rangle$, and they are closest to these three points.

\section{Comparison with optimal approximation based on trace distance}\label{Q2}
The different choices of distance measure and set of available quantum states will have certain influence on the optimal convex approximation of quantum states. In Sec. \ref{Q1}, taking $B_{3}$ as an example, we addressed the general problem of approximating an unavailable qubit state through fidelity. In Ref. \cite{99}, based on the trace norm, Liang et al. showed the analytical solutions in some cases by the convex mixing of quantum states in the set $B_{3}$. In this section, we analyze the advantages and disadvantages of the optimal convex approximation under fidelity by comparison the optiaml states obtained with these two distance measures.

Any qubit quantum state $\rho$ can also be expressed as
\begin{equation}
\begin{aligned}
\rho=\begin{pmatrix} 1-a & k\sqrt{a(1-a)}\textup{e}^{-\textup{i}\phi} \\ k\sqrt{a(1-a)}\textup{e}^{\textup{i}\phi} & a \end{pmatrix},
\end{aligned}
\end{equation}
where $a\in[0,1]$, $\phi\in[0,2\pi]$ and $k\in[0,1]$. Let $u=k\sqrt{a(1-a)}\textup{cos}\phi$ and $v=k\sqrt{a(1-a)}\textup{sin}\phi$, in fact, $a=\frac{1-r_{z}}{2}$, $u=\frac{r_{x}}{2}$, $v=\frac{r_{y}}{2}$.

In quantum mechanics, the difference between quantum states can be reflected essentially by the difference between their eigenvalues. Any two qubit states $\rho^{1}$ and $\rho^{2}$ can be written in diagonalized form $\rho^{1}=\lambda_{+}^{1}|\alpha_{+}\rangle\langle\alpha_{+}|+\lambda_{-}^{1}|\alpha_{-}\rangle\langle\alpha_{-}|$ and $\rho^{2}=\lambda_{+}^{2}|\beta_{+}\rangle\langle\beta_{+}|+\lambda_{-}^{2}|\beta_{-}\rangle\langle\beta_{-}|$ respectively. Here $\{|\alpha_{+}\rangle,|\alpha_{-}\rangle\}$ and $\{|\beta_{+}\rangle,|\beta_{-}\rangle\}$ are the two bases for a two-dimensional Hilbert space, they can be transformed into each other by unitary operators. If $\lambda_{\pm}^{1}=\lambda_{\pm}^{2}$, they are equivalent. Therefore, the problem of comparing the optimal states can be transformed into comparing the difference between the eigenvalues of the optimal state and the target state based on these two distances. By calculating, the eigenvalues of any qubit state $\rho$ are
\begin{equation}\label{10}
\begin{aligned}
\lambda_{\pm}=\pm\frac{\sqrt{r_{x}^{2}+r_{y}^{2}+r_{z}^{2}}}{2}+\frac{1}{2}=\pm\frac{|\pmb{r}|}{2}+\frac{1}{2}.
\end{aligned}
\end{equation}
The eigenvalues of the convex combination $\sum_{i}p_{i}\rho_{i}$ of available states are
\begin{equation}\label{19}
\begin{aligned}
h_{\pm}=\pm\frac{1}{2}\sqrt{(p_{0}-p_{1})^{2}+(p_{2}-p_{3})^{2}+(p_{4}-p_{5})^{2}}+\frac{1}{2}.
\end{aligned}
\end{equation}

We construct the difference function that quantifying the distance between the eigenvalues of the optimal state and the target state as
\begin{equation}\label{20}
\begin{aligned}
g=|h_{+}-\lambda_{+}|+|h_{-}-\lambda_{-}|.
\end{aligned}
\end{equation}

Based on the trace norm, the optimal convex approximation \cite{99} of $\rho$ with respect to $B_{3}$ is defined as $D_{B_{3}}(\rho)=\textup{min}\{||\rho-\sum_{i}p_{i}'\rho_{i}||_{1}\}$, where $\rho_{i}=|i\rangle\langle i|$, $p_{i}'\geq 0$, and $\sum_{i}p_{i}'=1$, the minimum is taken over all possible probability distributions $\{p_{i}'\}$. Let $\rho^{\textup{opt}'}$ denote the corresponding optimal state such that $D_{B_{3}}(\rho)=||\rho-\rho^{\textup{opt}'}||_{1}$. In the set $S1$, it is obvious that the value of difference function is zero under these two measures.

In Ref. \cite{99}, when only three of the probabilities $\{p_{i}'\}$ are nonzero, the probabilities of optimal state $\rho^{\textup{opt}'}$ are
\begin{equation}\label{21}
\begin{aligned}
&p_{0}'=1-\frac{4a}{3}-\frac{2u}{3}-\frac{2v}{3}=\frac{1}{3}+\frac{2r_{z}-r_{x}-r_{y}}{3},\\
&p_{2}'=\frac{2a}{3}+\frac{4u}{3}-\frac{2v}{3}=\frac{1}{3}+\frac{-r_{z}+2r_{x}-r_{y}}{3},\\
&p_{4}'=\frac{2a}{3}-\frac{2u}{3}+\frac{4v}{3}=\frac{1}{3}+\frac{-r_{z}-r_{x}+2r_{y}}{3},\\
&p_{1}'=p_{3}'=p_{5}'=0,
\end{aligned}
\end{equation}
the value range of $(r_{x},r_{y},r_{z})$ is the set $S2'=\{(r_{x},r_{y},r_{z})|r_{x}+r_{y}\leq 1+2r_{z},r_{y}+r_{z}\leq 1+2r_{x},r_{x}+r_{z}\leq 1+2r_{y}\}\setminus S1$. Our result, when only three values of the probabilities $\{p_{i}\}$ are nonzero, the coefficients of optimal state $\rho^{\textup{opt}}$ are the equation $(\ref{3})$, the value range of $(r_{x},r_{y},r_{z})$ is the set $S2$. It is apparent that $S2\subset S2'$. For the state $\rho$ belonging to set $S2$, we have the following conclusion.

$Proposition~1.$ In the set $S2$, the optimal state obtained by using the fidelity as a measure is closer to the expected state.

$Proof.$ In the set $S2$, according to the equations (\ref{19}) and (\ref{21}), we obtain the eigenvalues of $\rho^{\textup{opt}'}$ as
\begin{equation}\label{22}
\begin{aligned}
h_{\pm}^{1'}=\pm\frac{1}{2}\sqrt{\frac{1+3|\pmb{r}|^{2}-r^{2}}{3}}+\frac{1}{2},
\end{aligned}
\end{equation}
where $r=r_{x}+r_{y}+r_{z}$. Due to the equations (\ref{3}) and (\ref{19}), the eigenvalues of $\rho^{\textup{opt}}$ are
\begin{equation}\label{23}
\begin{aligned}
h_{\pm}^{1}=\pm\frac{1}{2}\sqrt{\frac{1+2|\pmb{r}|^{2}-r^{2}}{3-r^{2}}}+\frac{1}{2}.
\end{aligned}
\end{equation}
Let us to compare the difference functions. By using the results (\ref{10}) and (\ref{22}), the difference function between optimal state based on the trace distance and target state is
\begin{equation}\label{14}
\begin{aligned}
g^{1'}=|h_{+}^{1'}-\lambda_{+}|+|h_{-}^{1'}-\lambda_{-}|=|\sqrt{\frac{1+3|\pmb{r}|^{2}-r^{2}}{3}}-|\pmb{r}||.
\end{aligned}
\end{equation}
By the result (\ref{23}), it is not difficult to get the difference function between optimal state based on the fidelity and target state
\begin{equation}\label{15}
\begin{aligned}
g^{1}=|h_{+}^{1}-\lambda_{+}|+|h_{-}^{1}-\lambda_{-}|=|\sqrt{\frac{1+2|\pmb{r}|^{2}-r^{2}}{3-r^{2}}}-|\pmb{r}||.
\end{aligned}
\end{equation}

We have $g^{1'}-g^{1}\geq 0$. For the detailed calculation please see the Appendix \ref{B}. That is, the optimal state obtained by using fidelity is better than one obtained by using trace norm in the value range $S2$. This completes the proof of the proposition.

When only two values of the probabilities $\{p_{i}'\}$ are nonzero, the value ranges obtained by Ref. \cite{99} are the set $S3'=\{(r_{x},r_{y},r_{z})|1-r_{z}<r_{x}+r_{y}\leq 1+2r_{z},r_{y}+r_{z}\leq 1+2r_{x},r_{x}+r_{z}> 1+2r_{y}\}$, $S4'=\{(r_{x},r_{y},r_{z})|1-r_{z}<r_{x}+r_{y}\leq 1+2r_{z},r_{y}+r_{z}> 1+2r_{x},r_{x}+r_{z}\leq 1+2r_{y}\}$ and $S5'=\{(r_{x},r_{y},r_{z})|r_{x}+r_{y}> 1+2r_{z}\}$. In this case, we have the following conclusion.

$Proposition~2.$ In the set $S3'$, $S4'$ and $S5'$, the optimal state based on the fidelity is closer to the desired state.

$Proof.$ First, we consider the case in the value range $S3'$. At this case, only $p_{0}'$ and $p_{2}'$ are nonzero. The probabilities \cite{99} of the optimal state $\rho^{\textup{opt}'}$ are
\begin{equation}
\begin{aligned}
&p_{0}'=1-a-u=\frac{1}{2}-\frac{r_{x}-r_{z}}{2},\\
&p_{2}'=a+u=\frac{1}{2}+\frac{r_{x}-r_{z}}{2},\\
&p_{1}'=p_{3}'=p_{4}'=p_{5}'=0.
\end{aligned}
\end{equation}
And according to the equation (\ref{19}), the eigenvalues of $\rho^{\textup{opt}'}$ are
\begin{equation}\label{26}
\begin{aligned}
h_{\pm}^{2'}=\pm\frac{1}{2}\sqrt{\frac{1+2(r_{x}^{2}+r_{z}^{2})-(r_{x}+r_{z})^{2}}{2}}+\frac{1}{2}.
\end{aligned}
\end{equation}
While, when only two probabilities $p_{0}$ and $p_{2}$ are nonzero, the value range is the set $S3$. The eigenvalues of $\rho^{\textup{opt}}$ are
\begin{equation}\label{27}
\begin{aligned}
h_{\pm}^{2}=\pm\frac{1}{2}\sqrt{\frac{1+r_{x}^{2}+r_{z}^{2}-r_{y}^{2}-(r_{x}+r_{z})^{2}}{2-(r_{x}+r_{z})^{2}-2r_{y}^{2}}}+\frac{1}{2}.
\end{aligned}
\end{equation}
Combining the equations (\ref{10}) and (\ref{26}), we gain the difference function between optimal state obtained by using the trace distance and target state
\begin{equation}\label{16}
\begin{aligned}
g^{2'}=|h_{+}^{2'}-\lambda_{+}|+|h_{-}^{2'}-\lambda_{-}|=|\sqrt{\frac{1+2(r_{x}^{2}+r_{z}^{2})-(r_{x}+r_{z})^{2}}{2}}-|\pmb{r}||.
\end{aligned}
\end{equation}
In the light of the equation (\ref{27}), the difference function between optimal state obtained by using fidelity and target state is
\begin{equation}\label{17}
\begin{aligned}
g^{2}=|h_{+}^{2}-\lambda_{+}|+|h_{-}^{2}-\lambda_{-}|=|\sqrt{\frac{1+r_{x}^{2}+r_{z}^{2}-r_{y}^{2}-(r_{x}+r_{z})^{2}}{2-(r_{x}+r_{z})^{2}-2r_{y}^{2}}}-|\pmb{r}||.
\end{aligned}
\end{equation}
It is easy to know that $S3'\subset S3$. We can prove $g^{2'}-g^{2}\geq 0$, for the details please refer to the Appendix \ref{C}. So, in the set $S3'$ the proposition is valid. The other two cases are same for only $p_{0}$, $p_{4}$ and $p_{2}$, $p_{4}$ being nonzero. The proof is completed.

The regions which have not been analyzed so far are $S2'\cap S3$, $S2'\cap S4$, and $S2'\cap S5$. In these cases, we have the following inferences. In the value range $S2'\cap S3$, when $r^{2}\geq (r_{x}+r_{z})^{2}+2r_{y}^{2}$, if $1-3|\pmb{r}|^{2}+r^{2}\leq 0$, then $g^{1'}-g^{2}\geq 0$, this means that using the fidelity as the measure is more advantageous, otherwise, cannot judge. When $r^{2}\leq (r_{x}+r_{z})^{2}+2r_{y}^{2}$, if $1-3|\pmb{r}|^{2}+r^{2}\geq 0$, then $g^{1'}-g^{2}\leq 0$, namely using the trace norm as the measure has superiority, otherwise, cannot judge. Please refer to the Appendix \ref{D} for the details.

Analogously, in the domain of definition $S2'\cap S4$, the difference function between the optimal state obtained by using fidelity and the expected state is
\begin{equation}
\begin{aligned}
g^{3}=|\sqrt{\frac{1+r_{y}^{2}+r_{z}^{2}-r_{x}^{2}-(r_{y}+r_{z})^{2}}{2-(r_{y}+r_{z})^{2}-2r_{x}^{2}}}-|\pmb{r}||.
\end{aligned}
\end{equation}
When $r^{2}\geq (r_{y}+r_{z})^{2}+2r_{x}^{2}$, if $1-3|\pmb{r}|^{2}+r^{2}\leq 0$, then $g^{1'}-g^{3}\geq 0$, otherwise, cannot judge. When $r^{2}\leq (r_{y}+r_{z})^{2}+2r_{x}^{2}$, if $1-3|\pmb{r}|^{2}+r^{2}\geq 0$, then $g^{1'}-g^{3}\leq 0$, otherwise, cannot judge.

In the set $S2'\cap S5$, the difference function between the optimal state obtained by using the fidelity and the wished state is
\begin{equation}
\begin{aligned}
g^{4}=|\sqrt{\frac{1+r_{x}^{2}+r_{y}^{2}-r_{z}^{2}-(r_{x}+r_{y})^{2}}{2-(r_{x}+r_{y})^{2}-2r_{z}^{2}}}-|\pmb{r}||.
\end{aligned}
\end{equation}
When $r^{2}\geq (r_{x}+r_{y})^{2}+2r_{z}^{2}$, if $1-3|\pmb{r}|^{2}+r^{2}\leq 0$, then $g^{1'}-g^{4}\geq 0$, otherwise, cannot judge. When $r^{2}\leq (r_{x}+r_{y})^{2}+2r_{z}^{2}$, if $1-3|\pmb{r}|^{2}+r^{2}\geq 0$, then $g^{1'}-g^{4}\leq 0$, otherwise, cannot judge.

These results make us realize that it is meaningful to research the optimal convex approximation of desired state by taking advantage of the fidelity. This allows one to find the optimal state $\rho^{\textup{opt}}$ closer to the target state in some regions.

\section{The geometry of CR states}\label{Q3}
The CR states indicate that these states can be completely represented by the convex mixing of quantum states in usable set. They are most perfect in our research. Next, we will study their geometric properties. Because of the symmetry, we only consider the value range $r_{x},r_{y},r_{z}\geq 0$. In the absence of ambiguity, the following is no longer marked.

According to the solution of equation ($\ref{2}$), we know that if and only if $r_{x}+r_{y}+r_{z}\leq 1$, $D_{B_{3}}^{F}(\rho)=0$. In this case, the objective state $\rho$ is CR state about set $B_{3}$. From the view of the geometry, the region of CR states is the dark purple regions in FIG. 1. The region is called $\mathcal{R}_{CR}$. Its vertex coordinates are (1,0,0), (0,1,0), (0,0,1) and (0,0,0). The corresponding quantum states of these vertices are $|2\rangle=\frac{1}{\sqrt{2}}(|0\rangle+|1\rangle)$, $|4\rangle=\frac{1}{\sqrt{2}}(|0\rangle+\textup{i}|1\rangle)$, $|0\rangle$, $\frac{1}{2}(|0\rangle\langle 0|+|1\rangle\langle 1|)$. From FIG. 1, it can be seen that the convex combination of these points forms the region $\mathcal{R}_{CR}$. The all purple regions including dark and light purple represent all quantum states, which is called $\mathcal{R}_{Q}$. The relative volume of the CR states with respect to whole quantum states is
\begin{equation}
\begin{aligned}
\mathcal{V}_{\mathcal{R}_{CR}}/\mathcal{V}_{\mathcal{R}_{Q}}=\frac{1}{6}/\frac{\pi}{6}=\frac{1}{\pi}.
\end{aligned}
\end{equation}

\begin{figure}[h]
\begin{minipage}[t]{0.3\linewidth}
\centering
\includegraphics[width=1\textwidth]{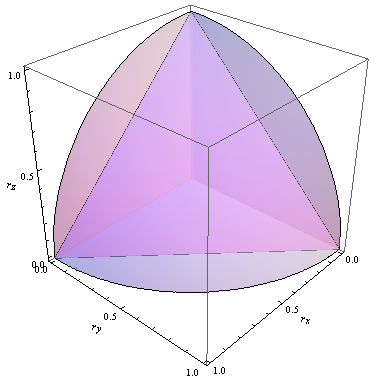}
\caption{The region $\mathcal{R}_{CR}$ is represented by dark purple. \label{11}}
\end{minipage}
\hfill
\begin{minipage}[t]{0.3\linewidth}
\centering
\includegraphics[width=1\textwidth]{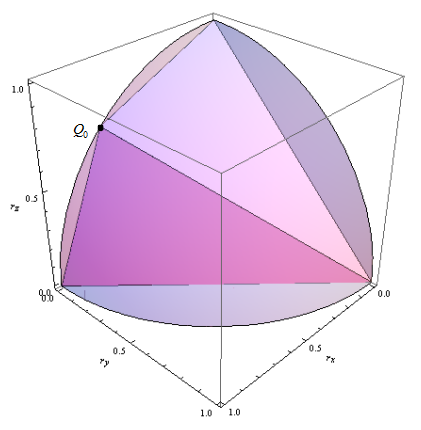}
\caption{The region $\mathcal{R}_{CR}^{\alpha_{0}}$ is expressed by dark purple. \label{12}}
\end{minipage}
\hfill
\begin{minipage}[t]{0.3\linewidth}
\centering
\includegraphics[width=1\textwidth]{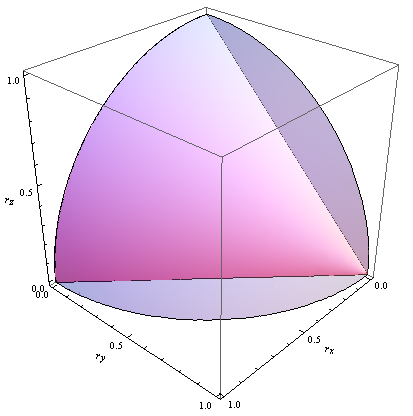}
\caption{The region $\mathcal{R}_{CR}^{\alpha}$ is denoted by dark purple. \label{13}}
\end{minipage}
\end{figure}

We can also consider other available states to expand the volume of the CR states. A real quantum logic gate is of the form either \cite{101}
\begin{equation}
\begin{aligned}
U_{\alpha}=\begin{pmatrix} \textup{cos}\alpha & \textup{sin}\alpha \\ \textup{sin}\alpha & -\textup{cos}\alpha \end{pmatrix}
~~\textup{or}~~
V_{\gamma}=\begin{pmatrix} \textup{cos}\gamma & -\textup{sin}\gamma \\ \textup{sin}\gamma & \textup{cos}\gamma \end{pmatrix}.
\end{aligned}
\end{equation}
The eigenvectors of $U_{\alpha}$ are $|\phi_{1}^{\alpha}\rangle=\textup{cos}\frac{\alpha}{2}|0\rangle+\textup{sin}\frac{\alpha}{2}|1\rangle$ and $|\phi_{2}^{\alpha}\rangle=\textup{sin}\frac{\alpha}{2}|0\rangle-\textup{cos}\frac{\alpha}{2}|1\rangle$. It is not difficult to find that $U_{\alpha}$ can be reduced to the $Z$ gate ($\sigma_{z}$), Hadamard gate, and $X$ gate ($\sigma_{x}$) in quantum information processing, when $\alpha$ is equal to 0, $\frac{\pi}{4}$, and $\frac{\pi}{2}$, respectively. The vectors $|4\rangle=\frac{1}{\sqrt{2}}(|0\rangle+\textup{i}|1\rangle)$ and $|5\rangle=\frac{1}{\sqrt{2}}(|0\rangle-\textup{i}|1\rangle)$ are the eigenvectors of $V_{\gamma}$ ($\gamma\neq 0,\pi$), which are also the eigenvectors of $Y$ gate ($\sigma_{y}$). Now, we consider a new set
\begin{equation}
\begin{aligned}
B_{3}^{\alpha_{0}}=B_{3}\cup\{|\phi_{1}^{\alpha_{0}}\rangle=\textup{cos}\frac{\alpha_{0}}{2}|0\rangle+\textup{sin}\frac{\alpha_{0}}{2}|1\rangle,
|\phi_{2}^{\alpha_{0}}\rangle=\textup{sin}\frac{\alpha_{0}}{2}|0\rangle-\textup{cos}\frac{\alpha_{0}}{2}|1\rangle\}.
\end{aligned}
\end{equation}
Here $\textup{cos}\frac{\alpha_{0}}{2}=\sqrt{\frac{2+\sqrt{2}}{4}}$, $\textup{sin}\frac{\alpha_{0}}{2}=\sqrt{\frac{2-\sqrt{2}}{4}}$, the quantum state $|\phi_{1}^{\alpha_{0}}\rangle$ is represented by the point $Q_{0}=\{\frac{\sqrt{2}}{2},0,\frac{\sqrt{2}}{2}\}$ in FIG. 2. It is evident that the convex combination of this usable states $|0\rangle$, $|2\rangle$, $|4\rangle$, $|\phi_{1}^{\alpha_{0}}\rangle$ and $I/2$ is the dark purple region in FIG. 2, this region is called $\mathcal{R}_{CR}^{\alpha_{0}}$. We come to the following conclusion.

$Proposition~3.$ The quantum state $\rho$ belongs to the region $\mathcal{R}_{CR}^{\alpha_{0}}$ if and only if $D_{B_{3}^{\alpha_{0}}}^{F}(\rho)=0$.

$Proof.$ Let $Q_{\rho}=(r_{x},r_{y},r_{z})$ be the corresponding point of quantum state $\rho$. For convenience, $\rho_{i}$ for $i=0,1,\ldots,4$ express $|0\rangle$, $|2\rangle$, $|4\rangle$, $|\phi_{1}^{\alpha_{0}}\rangle$ and $I/2$ respectively. $Q_{\rho_{i}}=(r_{x}^{i},r_{y}^{i},r_{z}^{i})$ denotes the point of quantum state $\rho_{i}$ in Bloch sphere.

First, we show that the proposition is valid for $Q_{\rho}\in \mathcal{R}_{CR}^{\alpha_{0}}$. From the characterization of convex combination, it is obvious that $Q_{\rho}$ can be linearly represented by the vertices of $\mathcal{R}_{CR}^{\alpha_{0}}$. More specifically, there is a set of weights $\{q_{i}\}$ with $q_{i}\geq 0$ and $\sum_{i}q_{i}=1$, such that $Q_{\rho}=\sum_{i}q_{i}Q_{\rho_{i}}$. Naturally, $r_{x}=\sum_{i}q_{i}r_{x}^{i}$. It is easy to know
\begin{equation*}
\begin{aligned}
\textup{Tr}(\rho\sigma_{x})=\sum_{i}q_{i}\textup{Tr}(\rho_{i}\sigma_{x})=\textup{Tr}(\sum_{i}q_{i}\rho_{i}\sigma_{x}).
\end{aligned}
\end{equation*}
It has the same form in the y-axis and the z-axis. This implies that $\rho=\sum_{i}q_{i}\rho_{i}$. And due to $I/2=\frac{1}{2}(|0\rangle\langle 0|+|1\rangle\langle 1|)$, the quantum state $\rho$ can be expressed by a convex combination of the states in $B_{3}^{\alpha_{0}}$. So, we have $D_{B_{3}^{\alpha_{0}}}^{F}(\rho)=0$.

Second, the reverse is still valid. This completes the proof of the proposition.

Therefore, the relative volume of the CR states under $B_{3}^{\alpha_{0}}$ with regard to all quantum states is
\begin{equation}
\begin{aligned}
\mathcal{V}_{\mathcal{R}_{CR}^{\alpha_{0}}}/\mathcal{V}_{\mathcal{R}_{Q}}=\frac{\sqrt{2}}{6}/\frac{\pi}{6}=\frac{\sqrt{2}}{\pi}.
\end{aligned}
\end{equation}

Next we discuss the available set
\begin{equation}
\begin{aligned}
B^{\alpha}=\{|4\rangle,|5\rangle,|\phi^{\alpha}\rangle=\textup{cos}\frac{\alpha}{2}|0\rangle+\textup{sin}\frac{\alpha}{2}|1\rangle\}.
\end{aligned}
\end{equation}
where $\alpha$ is taken over all the value from 0 to $2\pi$. The state $|\phi^{\alpha}\rangle$ is expressed by $Q_{|\phi^{\alpha}\rangle}=(\textup{sin}\alpha,0,\textup{cos}\alpha)$. In FIG. 3, the dark purple region is called $\mathcal{R}_{CR}^{\alpha}$. As a matter of fact, it is obtained by the rotation of point $Q_{|\phi^{\alpha}\rangle}$ around the circumference of the x-z plane. Similarly, we draw the following conclusion.

$Proposition~4.$ The quantum state $\rho$ of the region $\mathcal{R}_{CR}^{\alpha}$ satisfy $D_{B^{\alpha}}^{F}(\rho)=0$, and vice versa.

The proof is the same as above. Let $\rho_{i}$ for $i=4,5,6$ show the quantum state $|4\rangle$, $|5\rangle$, $|\phi^{\alpha}\rangle$ respectively, $p_{i}$ is the corresponding weight. From our optimal approximation analysis, it is easy to get the following conclusion.

$Proposition~5.$ For a qubit state $\rho$, if and only if $(1-r_{y})^2\geq r_{x}^{2}+r_{z}^{2}$, we have $D_{B^{\alpha}}^{F}(\rho)=0$. Meanwhile, $\textup{sin}\alpha=\frac{r_{x}}{\sqrt{r_{x}^{2}+r_{z}^{2}}}$, the coefficients of optimal state are
\begin{equation}\label{18}
\begin{aligned}
&p_{4}=\frac{1}{2}(1+r_{y}-\sqrt{r_{x}^{2}+r_{z}^{2}}),\\
&p_{5}=\frac{1}{2}(1-r_{y}-\sqrt{r_{x}^{2}+r_{z}^{2}}),\\
&p_{6}=\sqrt{r_{x}^{2}+r_{z}^{2}}.
\end{aligned}
\end{equation}

$Proof.$ First part, for arbitrary quantum state $\rho$, due to the Proposition 4, we deduce that $D_{B^{\alpha}}^{F}(\rho)=0$ if and only if $(1-r_{y})^2\geq r_{x}^{2}+r_{z}^{2}$.

Furthermore, when $\rho\in \mathcal{R}_{CR}^{\alpha}$, there must be a number $\alpha$, such that the state $\rho$ lies in the plane consisting of $\rho_{j}$, $j=4,5,6$. In the meantime, we have $\frac{\textup{sin}\alpha}{\textup{cos}\alpha}=\frac{r_{x}}{r_{z}}$. It can be converted to $\textup{sin}\alpha=\frac{r_{x}}{\sqrt{r_{x}^{2}+r_{z}^{2}}}$. From this, there is a set of probabilities $\{p_{j}\}$ with $p_{j}\geq 0$ and $\sum_{j}p_{j}=1$ for $j=4,5,6$ that makes $\rho=\sum_{j}p_{j}\rho_{j}$. This means that $Q_{\rho}=(\textup{sin}\alpha\cdot p_{6},p_{4}-p_{5},\textup{cos}\alpha\cdot p_{6})$. Then, we gain
\begin{equation}
\begin{aligned}
\textup{sin}\alpha\cdot p_{6}=r_{x},~~p_{4}-p_{5}=r_{y},~~\textup{cos}\alpha\cdot p_{6}={r_{z}},~~p_{4}+p_{5}+p_{6}=1.
\end{aligned}
\end{equation}
By solving the above equation, we obtain the solution (\ref{18}). That is the end of the proof.

So, the relative volume of the CR states concerning the set $B^{\alpha}$ with respect to entire quantum states is
\begin{equation}
\begin{aligned}
\mathcal{V}_{\mathcal{R}_{CR}^{\alpha}}/\mathcal{V}_{\mathcal{R}_{Q}}=\frac{\pi}{12}/\frac{\pi}{6}=\frac{1}{2}.
\end{aligned}
\end{equation}

The above results show that the range of CR states about the set $B^{\alpha}$ is the largest under the known usable states. In Ref. \cite{101}, researchers studied the optimal convex approximation of the qubit state $\rho$ in reference to the eigenvectors of arbitrarily two or three real quantum logic gates. They considered the available state sets $K_{1}=\{|\phi_{1}^{\alpha}\rangle,|\phi_{2}^{\alpha}\rangle,|\phi_{1}^{\beta}\rangle,|\phi_{2}^{\beta}\rangle\}$, $K_{3}=\{|\phi_{1}^{\alpha}\rangle,|\phi_{2}^{\alpha}\rangle,|4\rangle,|5\rangle\}$ and $K=\{|\phi_{1}^{\alpha}\rangle,|\phi_{2}^{\alpha}\rangle,|\phi_{1}^{\beta}\rangle,|\phi_{2}^{\beta}\rangle,|4\rangle,|5\rangle\}$. By analyzing the geometric properties of CR states, we can quickly know the range of CR states for different feasible sets. The region of CR states about the set $K_{1}$ is a sector with 90 degree central angle. The convex combinations of the states in the set $K_{3}$ and $K$ are both the region $\mathcal{R}_{CR}^{\alpha}$. The proposition 5 indicates that we can use less resource to express the states in the region $\mathcal{R}_{CR}^{\alpha}$.

\section{Conclusion}\label{Q4}
In summary, we define the optimal convex approximation based on the fidelity. For the set of eigenstates of all Pauli matrices, we have obtained the explicit analytical solution for an arbitrary qubit state. The advantage of our results is that the eigenvalues of the optimal state based on the fidelity are closer to the eigenvalues of the target state over the eigenvalues of the optimal state based on the trace norm in many ranges. Apart from the eigenstates of the Pauli matrices, we also consider the eigenvectors of other real quantum logic gates. We analytically calculate the volumes of the expected CR states in regard to several sets of available states, and find the relationship between the selected available states and CR states. The associated volume element depends only on the coordinates of gainable states with respect to three axes $\sigma_{x}$, $\sigma_{y}$ and $\sigma_{z}$. Finally, we completely represent the desired states in the region $\mathcal{R}_{CR}^{\alpha}$ with fewer available states.

\begin{acknowledgments}
This work was funded by the National Natural Science Foundation of China under Grant No. 12071110,
the Hebei Natural Science Foundation of China under Grant No. A2020205014, the Science and Technology Project of Hebei Education Department under Grant Nos. ZD2020167 and ZD2021066, and the Graduate Student Innovation Funding Project of School of Mathematical Sciences of Hebei Normal University under Grant No. 2021sxbs002.
\end{acknowledgments}

\begin{appendix}
\section{the process of solving the equations (\ref{2})}\label{A}
Solve the following equations
\begin{subequations}\label{a}
\begin{align}
&\frac{\partial G}{\partial p_{0}}=\frac{(p_{0}-p_{1})\sqrt{1-|\pmb{r}|^{2}}}{2\sqrt{1-(p_{2}-p_{3})^{2}-(p_{4}-p_{5})^{2}-
(p_{0}-p_{1})^{2}}}-\frac{r_{z}}{2}-\lambda_{0}-\lambda=0, \label{a1}\\
&\frac{\partial G}{\partial p_{1}}=-\frac{(p_{0}-p_{1})\sqrt{1-|\pmb{r}|^{2}}}{2\sqrt{1-(p_{2}-p_{3})^{2}-(p_{4}-p_{5})^{2}-
(p_{0}-p_{1})^{2}}}+\frac{r_{z}}{2}-\lambda_{1}-\lambda=0, \label{a2}\\
&\frac{\partial G}{\partial p_{2}}=\frac{(p_{2}-p_{3})\sqrt{1-|\pmb{r}|^{2}}}{2\sqrt{1-(p_{2}-p_{3})^{2}-(p_{4}-p_{5})^{2}-
(p_{0}-p_{1})^{2}}}-\frac{r_{x}}{2}-\lambda_{2}-\lambda=0, \label{a3}\\
&\frac{\partial G}{\partial p_{3}}=-\frac{(p_{2}-p_{3})\sqrt{1-|\pmb{r}|^{2}}}{2\sqrt{1-(p_{2}-p_{3})^{2}-(p_{4}-p_{5})^{2}-
(p_{0}-p_{1})^{2}}}+\frac{r_{x}}{2}-\lambda_{3}-\lambda=0, \label{a4}\\
&\frac{\partial G}{\partial p_{4}}=\frac{(p_{4}-p_{5})\sqrt{1-|\pmb{r}|^{2}}}{2\sqrt{1-(p_{2}-p_{3})^{2}-(p_{4}-p_{5})^{2}-
(p_{0}-p_{1})^{2}}}-\frac{r_{y}}{2}-\lambda_{4}-\lambda=0, \label{a5}\\
&\frac{\partial G}{\partial p_{5}}=-\frac{(p_{4}-p_{5})\sqrt{1-|\pmb{r}|^{2}}}{2\sqrt{1-(p_{2}-p_{3})^{2}-(p_{4}-p_{5})^{2}-
(p_{0}-p_{1})^{2}}}+\frac{r_{y}}{2}-\lambda_{5}-\lambda=0, \label{a6}\\
&\frac{\partial G}{\partial \lambda}=\sum_{j}p_{j}-1=0,~\lambda_{i}p_{i}=0,~i=0,\cdots,5.
\end{align}
\end{subequations}

Step 1, it is easy to obtain
\begin{equation}
\begin{aligned}
&-\lambda_{0}-\lambda_{1}-2\lambda=0,\\
&-\lambda_{2}-\lambda_{3}-2\lambda=0,\\
&-\lambda_{4}-\lambda_{5}-2\lambda=0.\\
\end{aligned}
\end{equation}

Step 2, (V1) if $p_{0},p_{1}\neq 0$, or $p_{2},p_{3}\neq 0$, or $p_{4},p_{5}\neq 0$, or at least four elements of $\{p_{i}\}$ are nonzero, then we have $\lambda_{i},\lambda=0$, so the equation (\ref{a}) is reduced to
\begin{equation}\label{24}
\begin{aligned}
&\frac{(p_{0}-p_{1})\sqrt{1-|\pmb{r}|^{2}}}{2\sqrt{1-(p_{2}-p_{3})^{2}-(p_{4}-p_{5})^{2}-
(p_{0}-p_{1})^{2}}}-\frac{r_{z}}{2}=0,\\
&\frac{(p_{2}-p_{3})\sqrt{1-|\pmb{r}|^{2}}}{2\sqrt{1-(p_{2}-p_{3})^{2}-(p_{4}-p_{5})^{2}-
(p_{0}-p_{1})^{2}}}-\frac{r_{x}}{2}=0,\\
&\frac{(p_{4}-p_{5})\sqrt{1-|\pmb{r}|^{2}}}{2\sqrt{1-(p_{2}-p_{3})^{2}-(p_{4}-p_{5})^{2}-
(p_{0}-p_{1})^{2}}}-\frac{r_{y}}{2}=0.
\end{aligned}
\end{equation}
It infers
\begin{equation}\label{25}
\begin{aligned}
&(p_{0}-p_{1})^{2}=\frac{r_{z}^{2}[1-(p_{2}-p_{3})^{2}-(p_{4}-p_{5})^{2}]}{1-r_{x}^{2}-r_{y}^{2}},\\
&(p_{2}-p_{3})^{2}=\frac{r_{x}^{2}[1-(p_{0}-p_{1})^{2}-(p_{4}-p_{5})^{2}]}{1-r_{y}^{2}-r_{z}^{2}},\\
&(p_{4}-p_{5})^{2}=\frac{r_{y}^{2}[1-(p_{0}-p_{1})^{2}-(p_{2}-p_{3})^{2}]}{1-r_{x}^{2}-r_{z}^{2}}.
\end{aligned}
\end{equation}
According (\ref{24}) and (\ref{25}), we find
\begin{equation}\label{51}
\begin{aligned}
p_{0}-p_{1}=r_{z},~~p_{2}-p_{3}=r_{x},~~p_{4}-p_{5}=r_{y},~~\sum_{i}p_{i}=1.
\end{aligned}
\end{equation}
Then, the solution is
\begin{equation}
\begin{aligned}
&p_{0}=\frac{1}{2}-\frac{r_{x}}{2}-\frac{r_{y}}{2}+\frac{r_{z}}{2}-t_{1}-t_{2},\\
&p_{1}=\frac{1}{2}-\frac{r_{x}}{2}-\frac{r_{y}}{2}-\frac{r_{z}}{2}-t_{1}-t_{2},\\
&p_{2}=r_{x}+t_{2},\\
&p_{3}=t_{2},\\
&p_{4}=r_{y}+t_{1},\\
&p_{5}=t_{1},
\end{aligned}
\end{equation}
where $t_{1}$ and $t_{2}$ are arbitrary non-negative numbers such that $p_{i}\geq 0$. The constraint $p_{i}\geq 0$ with $i=0,1,\ldots,5$ is transformed to $1-r_{x}-r_{y}-r_{z}\geq 0$. Let $S1=\{(r_{x},r_{y},r_{z})|r_{x}+r_{y}+r_{z}\leq 1\}$. Due to (\ref{51}), it is easy to know that $\rho$ can be absolutely expressed by $\sum_{i}p_{i}\rho_{i}$. That is to say $D_{B_{3}}^{F}(\rho)=0$ in the set $S1$.

(V2) For $1-r_{x}-r_{y}-r_{z}< 0$, there are the following eight cases where three elements of $\{p_{i}\}$ are nonzero. (i) $p_{0}\neq 0,~p_{2}\neq 0,~p_{4}\neq 0$; (ii) $p_{0}\neq 0,~p_{2}\neq 0,~p_{5}\neq 0$; (iii) $p_{0}\neq 0,~p_{3}\neq 0,~p_{4}\neq 0$; (iv) $p_{0}\neq 0,~p_{3}\neq 0,~p_{5}\neq 0$; (v) $p_{1}\neq 0,~p_{2}\neq 0,~p_{4}\neq 0$; (vi) $p_{1}\neq 0,~p_{2}\neq 0,~p_{5}\neq 0$; (vii) $p_{1}\neq 0,~p_{3}\neq 0,~p_{4}\neq 0$; (viii) $p_{1}\neq 0,~p_{3}\neq 0,~p_{5}\neq 0$.

In the case (i), we have $\lambda_{0},\lambda_{2},\lambda_{4}=0$, $p_{1},p_{3},p_{5}=0$, $\lambda=-\frac{\lambda_{1}}{2}=-\frac{\lambda_{3}}{2}=-\frac{\lambda_{5}}{2}< 0$ then
\begin{equation}\label{28}
\begin{aligned}
&\frac{p_{0}\sqrt{1-|\pmb{r}|^{2}}}{2\sqrt{1-p_{2}^{2}-p_{4}^{2}-p_{0}^{2}}}-\frac{r_{z}}{2}-\lambda=0,\\
&\frac{p_{2}\sqrt{1-|\pmb{r}|^{2}}}{2\sqrt{1-p_{2}^{2}-p_{4}^{2}-p_{0}^{2}}}-\frac{r_{x}}{2}-\lambda=0,\\
&\frac{p_{4}\sqrt{1-|\pmb{r}|^{2}}}{2\sqrt{1-p_{2}^{2}-p_{4}^{2}-p_{0}^{2}}}-\frac{r_{y}}{2}-\lambda=0.
\end{aligned}
\end{equation}
It means
\begin{equation}\label{29}
\begin{aligned}
&p_{0}^{2}=\frac{(r_{z}+2\lambda)^{2}[1-p_{2}^{2}-p_{4}^{2}]}{1-|\pmb{r}|^{2}+(r_{z}+2\lambda)^{2}},\\
&p_{2}^{2}=\frac{(r_{x}+2\lambda)^{2}[1-p_{0}^{2}-p_{4}^{2}]}{1-|\pmb{r}|^{2}+(r_{x}+2\lambda)^{2}},\\
&p_{4}^{2}=\frac{(r_{y}+2\lambda)^{2}[1-p_{0}^{2}-p_{2}^{2}]}{1-|\pmb{r}|^{2}+(r_{y}+2\lambda)^{2}}.
\end{aligned}
\end{equation}
According (\ref{28}) and (\ref{29}), we obtain
\begin{equation}
\begin{aligned}
&p_{0}=\frac{(r_{z}+2\lambda)}{\sqrt{1-|\pmb{r}|^{2}+(r_{z}+2\lambda)^{2}+(r_{x}+2\lambda)^{2}+(r_{y}+2\lambda)^{2}}},\\
&p_{2}=\frac{(r_{x}+2\lambda)}{\sqrt{1-|\pmb{r}|^{2}+(r_{z}+2\lambda)^{2}+(r_{x}+2\lambda)^{2}+(r_{y}+2\lambda)^{2}}},\\
&p_{4}=\frac{(r_{y}+2\lambda)}{\sqrt{1-|\pmb{r}|^{2}+(r_{z}+2\lambda)^{2}+(r_{x}+2\lambda)^{2}+(r_{y}+2\lambda)^{2}}},\\
&\sum_{i}p_{i}=1.
\end{aligned}
\end{equation}
It implies $\lambda=\frac{1}{6}[\pm\sqrt{\frac{3-(r_{x}+r_{y}+r_{z})^{2}}{2}}-(r_{x}+r_{y}+r_{z})]$.

Because $(r_{z}+2\lambda)+(r_{x}+2\lambda)+(r_{y}+2\lambda)\geq 0$, we get $\lambda=\frac{1}{6}[\sqrt{\frac{3-(r_{x}+r_{y}+r_{z})^{2}}{2}}-(r_{x}+r_{y}+r_{z})]$. The constraint $\lambda<0$ can be converted to $1-r_{x}-r_{y}-r_{z}< 0$. So one has
\begin{equation}
\begin{aligned}
&p_{0}=\frac{1}{3}-\frac{\sqrt{2}(r_{x}+r_{y}-2r_{z})}{3\sqrt{3-(r_{x}+r_{y}+r_{z})^{2}}},\\
&p_{2}=\frac{1}{3}-\frac{\sqrt{2}(-2r_{x}+r_{y}+r_{z})}{3\sqrt{3-(r_{x}+r_{y}+r_{z})^{2}}},\\
&p_{4}=\frac{1}{3}-\frac{\sqrt{2}(r_{x}-2r_{y}+r_{z})}{3\sqrt{3-(r_{x}+r_{y}+r_{z})^{2}}},\\
&p_{1}=p_{3}=p_{5}=0.
\end{aligned}
\end{equation}
And due to $p_{0},p_{2},p_{4}> 0$, the above solutions should satisfy four constraints, $r_{x}+r_{y}< \sqrt{1-2r_{z}^{2}}+r_{z}$ or $r_{x}+r_{y}< 2r_{z}$, $r_{y}+r_{z}< \sqrt{1-2r_{x}^{2}}+r_{x}$ or $r_{y}+r_{z}< 2r_{x}$, $r_{x}+r_{z}< \sqrt{1-2r_{y}^{2}}+r_{y}$ or $r_{x}+r_{z}< 2r_{y}$, and $r_{x}+r_{y}+r_{z}>1$.

Let $S2=\{(r_{x},r_{y},r_{z})|r_{x}+r_{y}< \sqrt{1-2r_{z}^{2}}+r_{z}~\textup{or}~r_{x}+r_{y}<2r_{z},r_{y}+r_{z}< \sqrt{1-2r_{x}^{2}}+r_{x}~\textup{or}~r_{y}+r_{z}< 2r_{x},r_{x}+r_{z}< \sqrt{1-2r_{y}^{2}}+r_{y}~\textup{or}~r_{x}+r_{z}< 2r_{y}\}\setminus S1$. In this range, it is not difficult to get
\begin{equation}
\begin{aligned}
\max F(\rho,\sum_{i}p_{i}\rho_{i})&=\frac{1}{2}+\frac{r_{x}+r_{y}+r_{z}}{6}+\frac{\sqrt{2[3-(r_{x}+r_{y}+r_{z})^{2}]}}{6}\\
&=\frac{1}{2}+\frac{r}{6}+\frac{\sqrt{2(3-r^{2})}}{6},
\end{aligned}
\end{equation}
where $r=r_{x}+r_{y}+r_{z}$.

Further, we have
\begin{equation}
\begin{aligned}
D_{B_{3}}^{F}(\rho)=\frac{1}{2}-\frac{r}{6}-\frac{\sqrt{2(3-r^{2})}}{6}.
\end{aligned}
\end{equation}

In the case (ii), we have $\lambda_{0},\lambda_{2},\lambda_{5}=0$, $p_{1},p_{3},p_{4}=0$, $\lambda=-\frac{\lambda_{1}}{2}=-\frac{\lambda_{3}}{2}=-\frac{\lambda_{4}}{2}< 0$ then
\begin{subequations}
\begin{align}
&\frac{p_{0}\sqrt{1-|\pmb{r}|^{2}}}{2\sqrt{1-p_{2}^{2}-p_{5}^{2}-p_{0}^{2}}}-\frac{r_{z}}{2}-\lambda=0, \label{b1}\\
&\frac{p_{2}\sqrt{1-|\pmb{r}|^{2}}}{2\sqrt{1-p_{2}^{2}-p_{5}^{2}-p_{0}^{2}}}-\frac{r_{x}}{2}-\lambda=0, \label{b2}\\
&\frac{p_{5}\sqrt{1-|\pmb{r}|^{2}}}{2\sqrt{1-p_{2}^{2}-p_{5}^{2}-p_{0}^{2}}}+\frac{r_{y}}{2}-\lambda=0. \label{b3}
\end{align}
\end{subequations}
According (\ref{b3}), we can know that $2\lambda-r_{y}\geq 0$. Clearly, $2\lambda-r_{y}<0$. So there is a contradiction.

By the same method as above, it is easy to obtain that there are no solution in the other six cases.

(V3) Now consider the cases that only two elements of $\{p_{i}\}$ are nonzero in the rest region.

$(1')$ For $p_{0},p_{3}\neq 0$, $\lambda_{0},\lambda_{3}=0$, $\lambda=-\frac{\lambda_{1}}{2}=-\frac{\lambda_{2}}{2}=-\frac{\lambda_{4}+\lambda_{5}}{2}<0$, then
\begin{subequations}
\begin{align}
&\frac{p_{0}\sqrt{1-|\pmb{r}|^{2}}}{2\sqrt{1-p_{3}^{2}-p_{0}^{3}}}-\frac{r_{z}}{2}-\lambda=0, \label{c1}\\
&\frac{p_{3}\sqrt{1-|\pmb{r}|^{2}}}{2\sqrt{1-p_{3}^{2}-p_{0}^{3}}}+\frac{r_{x}}{2}-\lambda=0. \label{c2}
\end{align}
\end{subequations}
The equation (\ref{c2}) implies $2\lambda-r_{x}\geq 0$. As we know $2\lambda-r_{x}< 0$. Hence there is a contradiction.

Similarly, in the cases $p_{0},p_{5}\neq 0$; $p_{1},p_{2}\neq 0$; $p_{1},p_{3}\neq 0$; $p_{1},p_{4}\neq 0$; $p_{1},p_{5}\neq 0$; $p_{2},p_{5}\neq 0$; $p_{3},p_{4}\neq 0$; $p_{3},p_{5}\neq 0$, there are no solution.

$(2')$ For $p_{0},p_{2}\neq 0$, $\lambda_{0},\lambda_{2}=0$, $\lambda=-\frac{\lambda_{1}}{2}=-\frac{\lambda_{3}}{2}=-\frac{\lambda_{4}+\lambda_{5}}{2}<0$, then
\begin{equation}\label{36}
\begin{aligned}
&\frac{p_{0}\sqrt{1-|\pmb{r}|^{2}}}{2\sqrt{1-p_{2}^{2}-p_{0}^{2}}}-\frac{r_{z}}{2}-\lambda=0,\\
&\frac{p_{2}\sqrt{1-|\pmb{r}|^{2}}}{2\sqrt{1-p_{2}^{2}-p_{0}^{2}}}-\frac{r_{x}}{2}-\lambda=0.
\end{aligned}
\end{equation}
It signifies
\begin{equation}\label{37}
\begin{aligned}
&p_{0}^{2}=\frac{(r_{z}+2\lambda)^{2}[1-p_{2}^{2}]}{1-|\pmb{r}|^{2}+(r_{z}+2\lambda)^{2}},\\
&p_{2}^{2}=\frac{(r_{x}+2\lambda)^{2}[1-p_{0}^{2}]}{1-|\pmb{r}|^{2}+(r_{x}+2\lambda)^{2}}.
\end{aligned}
\end{equation}
According (\ref{36}) and (\ref{37}), we deduce
\begin{equation}
\begin{aligned}
&p_{0}=\frac{(r_{z}+2\lambda)}{\sqrt{1-|\pmb{r}|^{2}+(r_{z}+2\lambda)^{2}+(r_{x}+2\lambda)^{2}}},\\
&p_{2}=\frac{(r_{x}+2\lambda)}{\sqrt{1-|\pmb{r}|^{2}+(r_{z}+2\lambda)^{2}+(r_{x}+2\lambda)^{2}}},\\
&\sum_{i}p_{i}=1.
\end{aligned}
\end{equation}
Thus, we obtain $\lambda=\frac{1}{4}[\pm\sqrt{2-(r_{x}+r_{z})^{2}-2r_{y}^{2}}-(r_{x}+r_{z})]$.

Because $(r_{z}+2\lambda)+(r_{x}+2\lambda)\geq 0$, we choose
$\lambda=\frac{1}{4}[\sqrt{2-(r_{x}+r_{z})^{2}-2r_{y}^{2}}-(r_{x}+r_{z})]$. Since $\lambda<0$, we have $1-2r_{x}r_{z}<r_{x}^{2}+r_{y}^{2}+r_{z}^{2}$ which can hold. It is easy to obtain
\begin{equation}
\begin{aligned}
&p_{0}=\frac{1}{2}-\frac{r_{x}-r_{z}}{2\sqrt{2-(r_{x}+r_{z})^{2}-2r_{y}^{2}}},\\
&p_{2}=\frac{1}{2}-\frac{-r_{x}+r_{z}}{2\sqrt{2-(r_{x}+r_{z})^{2}-2r_{y}^{2}}},\\
&p_{1}=p_{3}=p_{4}=p_{5}=0,
\end{aligned}
\end{equation}
where $r_{x}$, $r_{y}$, $r_{z}$ belong to the set $S3=\{(r_{x},r_{y},r_{z})|r_{x}+r_{z}>\sqrt{1-r_{y}^{2}}\}\setminus S1\cup S2$.

In this case, we have
\begin{equation}
\begin{aligned}
\max F(\rho,\sum_{i}p_{i}\rho_{i})&=\frac{1}{2}+\frac{\sqrt{2-(r_{x}+r_{z})^{2}-2r_{y}^{2}}+(r_{x}+r_{z})}{4}.
\end{aligned}
\end{equation}
Further, the optimal convex approximation of quantum state $\rho$ is
\begin{equation}
\begin{aligned}
D_{B_{3}}^{F}(\rho)=\frac{1}{2}-\frac{\sqrt{2-(r_{x}+r_{z})^{2}-2r_{y}^{2}}+(r_{x}+r_{z})}{4}.
\end{aligned}
\end{equation}

$(3')$ Analogously, for $p_{0},p_{4}\neq 0$, we get
\begin{equation}
\begin{aligned}
&p_{0}=\frac{1}{2}-\frac{r_{y}-r_{z}}{2\sqrt{2-(r_{y}+r_{z})^{2}-2r_{x}^{2}}},\\
&p_{4}=\frac{1}{2}-\frac{-r_{y}+r_{z}}{2\sqrt{2-(r_{y}+r_{z})^{2}-2r_{x}^{2}}},\\
&p_{1}=p_{2}=p_{3}=p_{5}=0.
\end{aligned}
\end{equation}
Here $r_{x}$, $r_{y}$, $r_{z}$ belong to the set $S4=\{(r_{x},r_{y},r_{z})|r_{y}+r_{z}>\sqrt{1-r_{x}^{2}}\}\setminus S1\cup S2$.

From this, one gets
\begin{equation}
\begin{aligned}
\max F(\rho,\sum_{i}p_{i}\rho_{i})&=\frac{1}{2}+\frac{\sqrt{2-(r_{y}+r_{z})^{2}-2r_{x}^{2}}+(r_{y}+r_{z})}{4}.
\end{aligned}
\end{equation}
Hence, we have
\begin{equation}
\begin{aligned}
D_{B_{3}}^{F}(\rho)=\frac{1}{2}-\frac{\sqrt{2-(r_{y}+r_{z})^{2}-2r_{x}^{2}}+(r_{y}+r_{z})}{4}.
\end{aligned}
\end{equation}

$(4')$ For the case $p_{2},p_{4}\neq 0$, similar with $(3')$ we have
\begin{equation}
\begin{aligned}
&p_{2}=\frac{1}{2}-\frac{r_{y}-r_{x}}{2\sqrt{2-(r_{y}+r_{x})^{2}-2r_{z}^{2}}},\\
&p_{4}=\frac{1}{2}-\frac{-r_{y}+r_{x}}{2\sqrt{2-(r_{y}+r_{x})^{2}-2r_{z}^{2}}},\\
&p_{0}=p_{1}=p_{3}=p_{5}=0,
\end{aligned}
\end{equation}
where $r_{x}$, $r_{y}$, $r_{z}$ belong to the set $S5=\{(r_{x},r_{y},r_{z})|r_{x}+r_{y}>\sqrt{1-r_{z}^{2}}\}\setminus S1\cup S2$.

As a result,
\begin{equation}
\begin{aligned}
\max F(\rho,\sum_{i}p_{i}\rho_{i})&=\frac{1}{2}+\frac{\sqrt{2-(r_{x}+r_{y})^{2}-2r_{z}^{2}}+(r_{x}+r_{y})}{4}.
\end{aligned}
\end{equation}
Thus, we obtain
\begin{equation}
\begin{aligned}
D_{B_{3}}^{F}(\rho)=\frac{1}{2}-\frac{\sqrt{2-(r_{x}+r_{y})^{2}-2r_{z}^{2}}+(r_{x}+r_{y})}{4}.
\end{aligned}
\end{equation}

\section{the comparison of two difference functions in the set \textit{S}2}\label{B}
It is not difficult to find $S2\cap S2'=S2$. In the set $S2$, the range of $r^{2}$ is $(1,3]$ because of $r>1$ and $r^{2}\leq[3\sqrt{\frac{|\pmb{r}|^{2}}{3}}]^{2}=3|\pmb{r}|^{2}$. The equations (\ref{14}) and (\ref{15}) can be simplified as $g^{1'}=|\pmb{r}|-\sqrt{\frac{1+3|\pmb{r}|^{2}-r^{2}}{3}}$, $g^{1}=|\pmb{r}|-\sqrt{\frac{1+2|\pmb{r}|^{2}-r^{2}}{3-r^{2}}}$. Thereby,
\begin{equation}
\begin{aligned}
g^{1'}-g^{1}=\sqrt{\frac{1+2|\pmb{r}|^{2}-r^{2}}{3-r^{2}}}-\sqrt{\frac{1+3|\pmb{r}|^{2}-r^{2}}{3}}.
\end{aligned}
\end{equation}
Due to the monotonicity of the power function $x^{\frac{1}{2}}$ on the domain $x\geq 0$, it only needs to know
\begin{equation}
\begin{aligned}
\frac{1+2|\pmb{r}|^{2}-r^{2}}{3-r^{2}}-\frac{1+3|\pmb{r}|^{2}-r^{2}}{3}=\frac{(3|\pmb{r}|^{2}-r^{2})(r^{2}-1)}{3(3-r^{2})}\geq 0.
\end{aligned}
\end{equation}
So, we obtain $g^{1'}-g^{1}\geq 0$. That is to say that $g^{1}$ is not greater than $g^{1'}$ in the set $S2$.

\section{the comparison of two difference functions in the set \textit{S}3$'$}\label{C}
It is obvious that $S3\cap S3'=S3'$. Thus, in the set $S3'$, the point satisfies $(r_{x}+r_{z})^{2}+r_{y}^{2}>1$. Further, we have $(r_{x}+r_{z})^{2}+2r_{y}^{2}> 1$. We cancel out the absolute value, the equations (\ref{16}) and (\ref{17}) become $g^{2'}=|\pmb{r}|-\sqrt{\frac{1+2(r_{x}^{2}+r_{z}^{2})-(r_{x}+r_{z})^{2}}{2}}$, $g^{2}=|\pmb{r}|-\sqrt{\frac{1+r_{x}^{2}+r_{z}^{2}-r_{y}^{2}-(r_{x}+r_{z})^{2}}{2-(r_{x}+r_{z})^{2}-2r_{y}^{2}}}$. In order to compare $g^{2'}$ with $g^{2}$, we compute
\begin{equation}
\begin{aligned}
g^{2'}-g^{2}=\sqrt{\frac{1+r_{x}^{2}+r_{z}^{2}-r_{y}^{2}-(r_{x}+r_{z})^{2}}{2-(r_{x}+r_{z})^{2}-2r_{y}^{2}}}-\sqrt{\frac{1+2(r_{x}^{2}+r_{z}^{2})-(r_{x}+r_{z})^{2}}{2}}.
\end{aligned}
\end{equation}
It is not difficult to know
\begin{equation}
\begin{aligned}
\frac{1+r_{x}^{2}+r_{z}^{2}-r_{y}^{2}-(r_{x}+r_{z})^{2}}{2-(r_{x}+r_{z})^{2}-2r_{y}^{2}}-\frac{1+2(r_{x}^{2}+r_{z}^{2})-(r_{x}+r_{z})^{2}}{2}
=\frac{(r_{x}-r_{z})^{2}[(r_{x}+r_{z})^{2}+2r_{y}^{2}-1]}{2[2-(r_{x}+r_{z})^{2}-2r_{y}^{2}]}\geq 0.
\end{aligned}
\end{equation}
Thus, we get $g^{2'}-g^{2}\geq 0$. In other word, in the set $S3'$, $g^{2}$ is not greater than $g^{2'}$.

\section{the comparison of two difference functions in the intersection of \textit{S}2$'$ and \textit{S}3}\label{D}
In the range $S2'\cap S3$, we have $r^{2}>1$, $(r_{x}+r_{z})^{2}+2r_{y}^{2}>1$. It can be seen from the above, the difference functions based on the two measures are $g^{1'}=|\pmb{r}|-\sqrt{\frac{1+3|\pmb{r}|^{2}-r^{2}}{3}}$ and $g^{2}=|\pmb{r}|-\sqrt{\frac{1+r_{x}^{2}+r_{z}^{2}-r_{y}^{2}-(r_{x}+r_{z})^{2}}{2-(r_{x}+r_{z})^{2}-2r_{y}^{2}}}$. By this, we obtain
\begin{equation}
\begin{aligned}
g^{1'}-g^{2}=\sqrt{\frac{1+r_{x}^{2}+r_{z}^{2}-r_{y}^{2}-(r_{x}+r_{z})^{2}}{2-(r_{x}+r_{z})^{2}-2r_{y}^{2}}}-\sqrt{\frac{1+3|\pmb{r}|^{2}-r^{2}}{3}}.
\end{aligned}
\end{equation}

When $r^{2}\geq (r_{x}+r_{z})^{2}+2r_{y}^{2}$, one deduces
\begin{equation}
\begin{aligned}
&\quad\frac{1+r_{x}^{2}+r_{z}^{2}-r_{y}^{2}-(r_{x}+r_{z})^{2}}{2-(r_{x}+r_{z})^{2}-2r_{y}^{2}}-\frac{1+3|\pmb{r}|^{2}-r^{2}}{3}\\
&=\frac{[3(1-|\pmb{r}|^{2})+(r^{2}-1)][1-(r_{x}+r_{z})^{2}-2r_{y}^{2}]+(r^{2}-1)}{3[2-(r_{x}+r_{z})^{2}-2r_{y}^{2}]}\\
&\geq \frac{[1-3|\pmb{r}|^{2}+r^{2}][1-(r_{x}+r_{z})^{2}-2r_{y}^{2}]}{3[2-(r_{x}+r_{z})^{2}-2r_{y}^{2}]}.
\end{aligned}
\end{equation}
The inequality is obtained by narrowing $r^{2}$ to $(r_{x}+r_{z})^{2}+2r_{y}^{2}$. Due to $1-(r_{x}+r_{z})^{2}-2r_{y}^{2}<0$, if $1-3|\pmb{r}|^{2}+r^{2}\leq 0$, then $g^{1'}-g^{2}\geq 0$, otherwise, cannot judge.

When $r^{2}\leq (r_{x}+r_{z})^{2}+2r_{y}^{2}$, we have
\begin{equation}
\begin{aligned}
\frac{1+r_{x}^{2}+r_{z}^{2}-r_{y}^{2}-(r_{x}+r_{z})^{2}}{2-(r_{x}+r_{z})^{2}-2r_{y}^{2}}-\frac{1+3|\pmb{r}|^{2}-r^{2}}{3}\leq \frac{[1-3|\pmb{r}|^{2}+r^{2}][1-(r_{x}+r_{z})^{2}-2r_{y}^{2}]}{3[2-(r_{x}+r_{z})^{2}-2r_{y}^{2}]}.
\end{aligned}
\end{equation}
The inequality comes from amplifying $r^{2}$ to $(r_{x}+r_{z})^{2}+2r_{y}^{2}$. Because $1-(r_{x}+r_{z})^{2}-2r_{y}^{2}<0$, if $1-3|\pmb{r}|^{2}+r^{2}\geq 0$, then $g^{1'}-g^{2}\leq 0$, otherwise, cannot judge.

In the range $S2'\cap S4$ and $S2'\cap S5$, the results are similar.

\end{appendix}

\end{document}